\documentstyle[12pt]{article}
\advance\textheight by \topskip
\begin{document}

\rightline{SU-ITP-93-1}
\rightline{hep-th/9301082}
\rightline{January 1993}

\vspace{.8cm}
\begin{center}
{\large\bf  Evolution of Pure States into Mixed States  \\}

\vskip .9 cm

{\bf
Jun Liu
\footnote{E-mail address: junliu@scs.slac.stanford.edu} }
 \vskip 0.1cm
Physics Department, Stanford University,
Stanford    CA 94305
\end{center}
\vskip .6 cm
\centerline{\bf ABSTRACT}
\vspace{-0.7cm}
\begin{quotation}
In the formulation of Banks, Peskin and Susskind,
we show that one can construct evolution equations for
the quantum mechanical density matrix $\rho$ with
operators which do not commute with hamiltonian  which evolve
pure states into mixed states, preserve the normalization and
positivity of $\rho$  and conserve energy. Furthermore,
it seems to be
different from  a quantum mechanical system with random
sources.
\end{quotation}

\normalsize
\newpage

\section{Introduction }
It is difficult to construct a consistent theory that combine quantum
mechanics and general relativity.
This has lead to considerations that generalize either
quantum mechanics (QM) or general relativity or both.
In QM, a state is characterized by the ray of a vector $\psi$ in a Hilbert
space.
The Schroedinger equation which describe   the time evolution of a state
is a equation for the vector which characterize the state:
$$
\dot{\psi}=-{\rm i}H\psi.
$$
The rays of vectors in a Hilbert space are in 1-1 correspondence with the
positive hermitian matrices
which can be diagonalize by a unitary matrix into the following form
$$
\rho={\rm diag}. (1, 0, 0, ...).
$$
It is obvious that there are many hermitian and positive matrices
which do not belong to this class.
One possible generalization of QM is to characterize the states
by positive definite matrices (density matrices) instead of vectors
in the Hilbert space. In this case, the subset of states which have
the above correspondence to vectors (i.e., all the state of QM)
are called pure states and the other mixed states.
The Schroedinger equation can also be written as an equation
for the density matrix:
$$
\dot{\rho}=-{\rm i}[H,\rho].
$$
It evolves the pure states into pure states.
In his study of black-hole physics, Hawking proposed a generalization of QM
which allows evolution of pure states into mixed states.

Hawking's idea was closely scrutinized by
Banks, Peskin and Susskind (BPS) \cite{BPS} (see also \cite{gross,ellis}).
They studied
the possible evolution equation of $\rho$ which preserve the normalization,
linearity, hermiticity.  They found the
most general  equation
\begin{eqnarray}
\rho=-i[H,\rho]-1/2\sum h_{\alpha\beta}[Q^\beta Q^\alpha \rho+
\rho Q^\beta Q^\alpha-2Q^\alpha\rho Q^\beta].
\end{eqnarray}
The condition that $\rho$ remains positive  for general $n$-dimension
 is not straightforward.
BPS showed that  a positive $h$  is a sufficient condition. They
further showed that for real, symmetric, positive $h$, the  above evolution
equation describe a QM system with random sources:
$$
H=H_0+\sum j_\alpha(t)Q^\alpha.
$$
They then  argued that for a QM system with random sources it is impossible
to require energy-momentum conservation and locality at the same time.
 This severely
restricts the generalization of QM along this direction.

Recently Srednicki \cite{srednicki} investigated
the constraint on the evolution equation due to Lorentz invariance.
He concluded that it is difficult
to have a Lorentz covariant evolution equation if only $Q$'s which commute with
$H$ are used. This lead him to construct
 evolution equations with
 $Q$'s which do not commute with $H$.
He gave on such example. But in his example, a positive
 density matrix may develop into an indefinite one.

In this notes, we will study in detail the case of 2-dimension
where  sufficient and necessary condition for the positivity is
straightforward.
We will give an evolution equation using $Q$'s which do not commute with
$H$. Our example is an improvement of Srednicki's because in our example
 a positive density matrix will remain positive under the evolution.
Furthermore, despite of some similarities,
an evolution equation with an indefinite $h$ can not describe
 a QM system with a random source,
because such system requires $Q$ commute with $H$ if energy is conserved.
This result is first pointed out by BPS but the proof was not given
in \cite{BPS} and so we will give a proof in section 3.

\section{ 2 dimension}
In the 2 dimension, we can choose the identity matrix
$I$ and Pauli matrices $\{\vec{\sigma}\}$
as the basis for hermitian matrices.
We can write a normalized density matrix $\rho$ (${\rm Tr}\rho=1$) as
$$
\rho=1/2(I+\vec{\rho}\cdot\vec{\sigma}).
$$
Since we can always transform $\vec{\rho}\cdot\vec{\sigma}$ into
$|\vec{\rho}|\sigma_3$, it is obvious that $\rho\geq 0$ is equivalent to
$|\vec{\rho}|\leq 1$.
Now the master equation for $\rho$ can be written as an equation for
$\vec{\rho}$:
\begin{eqnarray}
\dot{\vec{\rho}}=-(S+A)\vec{\rho}+\vec{\beta},
\end{eqnarray}
where $S$ is a symmetric matrix,
\begin{eqnarray}
S=
\left(\begin{array}{ccc}2(h_{22}+h_{33}) & -h_{12}-h_{21} & -h_{13}-h_{31}\\
                      -h_{12}-h_{21} &  2(h_{11}+h_{33}) & -h_{23}-h_{32} \\
                      -h_{13}-h_{31} & -h_{23}-h_{32} & 2(h_{11}+h_{22})
     \end{array}
\right),
\end{eqnarray}
 $\vec{\beta}$ is a vector,
\begin{eqnarray}
\vec{\beta}=-{\rm i}
\left(\begin{array}{c}h_{23}-h_{32}\\ h_{13}-h_{31}\\
h_{12}-h_{21}
     \end{array}
\right),
\end{eqnarray}
and $A$ is an antisymmetric matrix which comes from
the first term in BPS equation, $-i[H,\rho]$.
This leads to
$$
\frac{\rm d}{\rm dt}(\vec{\rho})^2=2\vec{\rho}\cdot\dot{\vec{\rho}}
=-(\vec{\rho})^T(S+A)\vec{\rho}+\vec{\beta}\cdot\vec{\rho}
=-(\vec{\rho})^TS\vec{\rho}+\vec{\beta}\cdot\vec{\rho}.
$$
{}From the above equations, we see that $S$ is diagonalized and $\vec{\beta}=0$
when $h$ is diagonalized.
Thus it follows that the necessary and sufficient condition
for all $\rho$ to be positive is that
$$S\geq 0.$$
One  can     readily see that $S\geq 0$ when   $h\geq 0$,
which is the sufficient condition found by BPS.
But there are certainly  indefinite $h$ which  lead to
$S\geq 0$. This is true even after we impose the energy conservation
condition as we will show now.

Without the loss of
generality, we can take $H=\sigma_3$, and in this case, $H^2=I$,
the energy conservation reduce to $Tr(\sigma_3\rho)=0$.
This is equivalent to
$$
h_{13}=-h_{31}, h_{23}=-h_{32}, h_{11}+h_{22}=0.
$$
We can now construct examples using indefinite $h$ that conserve the energy,
preserve the positivity of density matrices.
For example, we can take
$h_{11}=-h_{22}=h_{33}=g$,
$h$ is  indefinite.
This gives the following equation in terms fermionic creation
and annihilation operators
$$
\dot{\rho}=[-{\rm i}[H^0,\rho]-2g(b^+\rho b^++b\rho b)]
+2g(b^+b\rho+\rho b^+b-2b^+b\rho b^+b).
$$
Without the second term, $2g(b^+b \rho+\rho b^+b-2b^+b\rho b^+b)$,
it is the example of Srednicki.
It fails to preserve the positivity.
With the second term,
the positivity is preserved.

When $h$ is non-negative, the system is just a QM system with a random
source, with the following hamiltonain,
$$
H_T=H+\sum_{\alpha}j_\alpha (t)Q^\alpha,
$$
where $j_\alpha(t)$ is a random source.
 This is showed by BPS and it provides a physical intuition
for working with this system.
Since $h_{22}=-1$, one would naturally ask if this can realized as
QM system with random source which has a non-hermitian Hamiltonian:
$$
H_T=H+{\rm i}j(t)Q.
$$
Since $H_T$ is not hermitian, the evolution of the total
system cannot be both unitary and we
will lose either hermiticity or normalization of $\rho_T$.

If we choose to preserve the hermiticity,
then we can use the following evolution equation
$$
\dot{\rho_T}=H_T\rho_T +\rho_T H_T.
$$
This does not lead to the BPS equation.
We can also choose to preserve the probability,
using the following equation
$$
\dot{\rho_T}={\rm i}[H_T,\rho_T].
$$
This evolution equation does not preserve the hermiticity of $\rho_T$,
but after the ensemble average, the hermiticity of
ensemble average density matrix $\bar{\rho_T}=\rho$ is recovered
and it leads to the BPS equation.
However, despite this formal similarity, the evolution equation with
indefinite $h$ seems  to be genuinely different from  the one
with non-negative $h$, as we will show in the next section.

 \section{n dimension}
We consider the case of dimension $n$ in this section.
We will prove that energy conservation
requires all the $Q_\alpha$ commute with $H$ if $h$
is real symmetric and positive.
This is first pointed out by BPS, but the proof was
not given in \cite{BPS}.  Here $\vec{\rho}$ is a $n^2-1$ vector
and the most general $H$ is
$$
H=\sum_a h_aQ^a_0,
$$
where $Q^a_0$ belong to the Cartan subalgebra of $SU(n^2-1)$.
In other word, $H$ is diagonal and $Q^a_0$'s are a basis
of hermitian diagonal matrices.
Since $H^k$ can be expressed in terms of $I$ and $Q^a_0$ for
arbitrary $k$, the energy conservation is equivalent to
$$
Tr(Q^a_0\rho(t))=0,
$$
for all the $Q^a_0$ belong to the Cartan subalgebra.
 We will
use $a$ (instead of $\alpha$) as the index to denote these matrices.
The above equation is equivalent to
$$
\dot{\rho}_a(t)=0.
$$
This is equivalent to $S_{a\alpha}=0$ for all $a$ and $\alpha$.
Now let us assume that $h_{\alpha\beta}$ is real, symmetric
and non-negative. Then we can diagonalize $h$:
$$h={\rm diag.}(h_1...h_{n^2-1}),$$
where $h_\alpha\geq0$.
We have
$$
\dot{\rho}=-1/2\sum_\gamma h_\gamma[Q^\gamma,[Q^\gamma,\rho]].
$$
So
$$
\dot{\rho}_a=-1/2 \sum_{\alpha\beta\gamma}h_\gamma f_{\gamma\beta a}
f_{\gamma\beta\alpha}\rho_\alpha,
$$
where $f_{\alpha\beta\gamma}$'s are structure coefficients of $SU(n^2-1)$.
That is
$$
S_{aa}=-1/2\sum_{\gamma,\beta}h_\gamma f^2_{\gamma\beta a}.
$$
 For every $Q^\gamma\not\in$Cartan subalgebra
 there exist
$\beta$ and $a$ such that
$f_{\gamma\beta a}\neq 0$. Since $h_\alpha>0$
 we have $S_{aa}>0$ if $Q^\gamma \not\in$ Cartan subalgebra.
That is, energy conservation  requires $Q$ commute with $H$.

Combined with the example of section 2, this result
indicate that the evolution equation with
indefinite $h$ is truly different from the ones with positive
$h$. In the latter case, energy conservation requires $Q$'s to be
conserved charges which is a very strong constraint. With indefinite
$h$, we can evade this constraint as the example in section 2.

 \section{Conclusion}

It seems that there is a very interesting class of evolution equations
which cannot be realized as a quantum mechanics system with random sources.
The condition for energy conservation on this class of
evolution is not as stringent, for example,
one can choose $Q$ which does not
commute with Hamiltonian $H$.

Srednicki \cite{srednicki} has argued that in quantum field theory
it is difficult to construct a Lorentz covariant evolution equation
with only $Q$'s which commute with $H$. He pointed
out that one possible way out is to use $Q$'s which do not commute with
$H$. In this case, one must make sure that the energy is conserved.
He gave one such  example, but the evolution equation leads to negative
$\rho$. Our example has no such flaw and is an improvement of
 Srednicki's result.

Generalization to filed theory example is under investigation. Hopefully,
the requirement of momentum conservation and locality
 will also become less stringent.

\section*{Acknowledgements}
The author  is grateful to L. Susskind for insightful discussions and
important suggestions and for comments on the manuscript.
He wish to thank L. Thorlacius  for useful discussions and
the  World Laboratory for a world lab scholarship.
He also would like to thank G. Lindblad for pointing out
mistakes in the earlier version of the paper.
\vskip 1cm


\end{document}